\documentclass[onecolumn,english,showpacs,preprintnumbers,amsmath,amssymb,floatfix]{revtex4}
\usepackage[T1]{fontenc}
\usepackage{color}
\usepackage{array}
\usepackage{amstext}
\usepackage{graphicx}
\usepackage{esint}

\makeatletter

\@ifundefined{textcolor}{}
{%
 \definecolor{BLACK}{gray}{0}
 \definecolor{WHITE}{gray}{1}
 \definecolor{RED}{rgb}{1,0,0}
 \definecolor{GREEN}{rgb}{0,1,0}
 \definecolor{BLUE}{rgb}{0,0,1}
 \definecolor{CYAN}{cmyk}{1,0,0,0}
 \definecolor{MAGENTA}{cmyk}{0,1,0,0}
 \definecolor{YELLOW}{cmyk}{0,0,1,0}
 }

\@ifundefined{definecolor}
 {\usepackage{color}}{}
\@ifundefined{definecolor}
 {\usepackage{color}}{}
\makeatother

\makeatother

\usepackage{babel}
\begin{document}

\title{Hadron energy spectrum in polarized top quark decays 
considering the effects of hadron and bottom quark masses}

\author{S. Mohammad Moosavi Nejad$^{a,b}$}
\email{mmoosavi@yazd.ac.ir}

\author{Mahboobe Balali$^a$}

\affiliation{$^{(a)}$Faculty of Physics, Yazd University, P.O. Box
89195-741, Yazd, Iran}

\affiliation{$^{(b)}$School of Particles and Accelerators,
Institute for Research in Fundamental Sciences (IPM), P.O.Box
19395-5531, Tehran, Iran}

\date{\today}

\begin{abstract}

We present the analytical expressions for the next-to-leading order
corrections to the partial decay width $t(\uparrow) \rightarrow bW^+$, followed by $b\rightarrow H_bX$,
for nonzero b-quark mass ($m_b\neq 0$) in the fixed-flavor-number scheme (FFNs).
To make the  predictions for the energy distribution of  outgoing hadrons $H_b$, as a
function of the normalized $H_b$-energy fraction $x_H$, we apply the general-mass variable-flavor-number scheme (GM-VFNs)
in a  specific helicity coordinate system where the polarization of top quark is evaluated relative to the b-quark momentum.
We also study the effects of gluon fragmentation and finite  hadron mass on the hadron energy spectrum so that hadron masses
are responsible for the low-$x_H$ threshold.
In order to describe both the b-quark and the gluon hadronizations in top decays we apply realistic and nonperturbative
fragmentation functions extracted through a global fit to $e^+e^-$ annihilation data from CERN LEP1 and SLAC SLC by relying on their universality and scaling violations.

\end{abstract}


\maketitle

\section{Introduction}
Ever since the top quark discovery in $p\bar{p}$ collisions in 1995 by the CDF \cite{Abe:1995hr}
and D0 experiments \cite{Abachi:1995iq}  at the Fermilab Tevatron, it
has been in or near the center of attention in high-energy physics. 
Its characteristics such as its mass $m_t$, total decay width $\Gamma_t$, branching fractions, and 
elements $V_{tq}$ of the Cabibbo-Kobayashi-Maskawa (CKM)
quark mixing matrix, have not yet been determined precisely. 
Its property which is likely most central in many aspects of top physics is its mass.
Since, for example, the top interacts
with the Higgs boson through the potential $y_t h\bar{t} t$ where $y_t$ is proportional to the top mass $m_t$, therefore 
the top mass has a fundamental role in the issue of stability of the Higgs potential (see Eq.~(\ref{hasan})).
Using the full sample of $p\bar{p}$ collision
data collected by the $D0$ experiment in the Tevatron Run II \cite{Abazov:2014dpa} at a
center-of-mass energy of $1.96$~TeV and for an integrated
luminosity of up to $8.7$~$fb^{-1}$ the mass of top quark is measured as $m_t=174.98\pm 0.76$~GeV.
Due to its remarkably large mass, the top couples strongly to 
the agents of electroweak symmetry breaking and this  makes it both an 
object of interest itself, and a tool to investigate
that mechanism in detail.\\
Among other things, the CERN LHC is 
a genuine top factory, in particular in Run II, producing about 90 million top quark pairs per year of running at 
 design energy $14$~TeV.
The existing and upcoming data  will allow us to study the top quark 
and its behavior in LHC collisions in great detail, if also the theoretical descriptions 
and simulations are of proportionate quality.\\
The top decay width itself is very difficult to determine in  hadron colliders, though a recent experimental inference 
of the width  was performed by D0 \cite{Abazov:2012vd}
and found  $\Gamma_t=2.00\pm 0.47$~GeV in the context of single top t-channel production, and CDF \cite{Aaltonen:2013kna} also
reported  $1.10<\Gamma_t<4.05$~GeV at the $68\%$ confidence level.
The top decay characteristics play an important role in studying
the top quark at colliders. 
Since, the width-to-mass ratio $\Gamma_t/m_t$ of the top quark is small enough then, for many purpose, 
the notion of top quark as a stable particle makes sense, so that 
its production and decay processes can be factorized through the narrow width approximation \cite{delDuca:2015gca}.
In fact, if it were not for the confinement of color, the top
could be considered as a free particle.
This property allows it to behave like a real particle and one can safely describe its decay in perturbative theory.\\
The top decay width is largely due to decays to a W-boson and a bottom quark (with the mass $m_b$), as it is represented in 
the element $|V_{tb}|\approx1$ of the CKM matrix \cite{Cabibbo:1963yz}.
Since $m_t>>m_W+m_b$, its width is sufficiently large to pre-empt top quark hadronization, 
then this rapid decay ($\tau_t\approx 0.5\times 10^{-24}$~s \cite{Chetyrkin:1999ju}) 
enables transmission of top quark spin information to final states.
Part of the top attractiveness  is its power to self-analyze its spin, through its purely left-handed 
Standard Model (SM) weak decay. 
The interplay between the top spin and its mass is of crucial importance in studying the SM.
The top quark polarization can be studied by the angular correlations 
 between the top quark spin and its decay product momenta,
 and these spin-momentum correlations will allow the detailed studies of the top decay mechanism.
In \cite{Nejad:2013fba}, we showed that these correlations depend on the choice of the possible helicity coordinate systems.\\
Since,  produced b-quarks hadronize before they decay, 
then each $b$-jet  contains a bottom flavored hadron
 which, most of the times, is a B-meson. They are identified by a displaced
decay vertex associated which charged lepton tracks.
In \cite{Corcella,Corcella:2001hz}, it is identified that the  hadronization of the b-quark
is the largest source of uncertainty in the measurement of the top mass at the LHC 
and the Tevatron.
These nonperturbative transitions are described by realistic, nonperturbative
fragmentation functions (FFs) that are usually obtained through a global fit to $e^+e^-$ data.
At LHC, the decay process $t\rightarrow BW^++X$ is of prime importance, and it
is an urgent task to predict its partial decay width as reliably as possible, specifically the distribution
in the scaled-energy of B-mesons ($x_B$) in the top quark rest frame is of particular interest.  
These $x_B$-distributions provide direct access to the B-meson FFs.
In \cite{Nejad:2013fba}, using the zero-mass variable-flavor-number scheme (ZM-VFNs)
in which the mass of b-quarks are set to zero at the parton-level,
we studied the NLO angular distribution of the 
scaled-energy of  B-hadrons through polarized top decays. For that, we calculated the polar angular correlation 
in the rest frame decay of a polarized top quark 
into a stable $W^+$-boson and the B-hadron, i.e. $t(\uparrow)\rightarrow W^++b(\rightarrow B+X)$.
 We analysed this correlation in a helicity 
coordinate system  where the event plane, including the top quark and its decay 
products, is defined in the $(x, z)$-plane with the Z-axes along the b-quark momentum.
Here, the top polarization vector was evaluated with respect to the b-momentum direction.\\
Here, using the same frame, we revisit B-hadron production from polarized top decays 
by working at NLO in the general-mass variable-flavor-number scheme (GM-VFNs), where
 b-quark masses are preserved from the beginning.
This makes the calculations more complicated.
Being manifestly based on Collin's QCD factorization theorem \cite{Collins:1998rz}
convenient for massive quarks, this 
factorization scheme allows us to resum the large logarithms 
in $m_b$, to retain the finite-$m_b$ effects and to preserve the universality of the FFs, 
whose scaling violations remain to be subject to DGLAP evolution \cite{dglap}.
In this way, it combines the virtues of the fixed-flavor-number scheme (FFNs) and the ZM-VFN scheme  and
also avoids their flaws. In fact, it is an elaborat tool for global analyses of experimental 
data on the inclusive production of heavy flavored hadrons, allowing one to transfer 
nonperturbative information on the hadronization of partons
from one type of experiment to another and from one 
energy scale $\mu_F$ to another, without the restriction $\mu_F>>m_b$ which is essential for the ZM-VFNs.
Our analysis is supposed to enhance our previous result \cite{Nejad:2013fba}
in the ZM-VFN scheme by retaining all nonlogarithmic $m_b$-terms.\\
Moreover, we also include finite-$m_B$ effects, 
which modify the relations between hadronic and partonic scaling variables and reduce the available phase space. 
However, due to the smallness $m_B/m_t$, we do not expect to measure these additional effects truly, except for certain 
corners the phase space. Their study is nevertheless necessary  to fully exploit the 
enormous statistics of the LHC data to be taken in the long run for a high precision determination of 
the top  properties. 
Studying top decays could be important to deepen our conception of the 
nonperturbative aspects of B-mesons  formation  and to test the universality and scaling violations of the B-meson FFs.

\section{Top quark in the Standard Model}
At first, we  briefly review the various interactions of the top quark field $t(x^\mu)$ in the SM Lagrangian; 
a topic needed for the calculation of top decay widths.\\
The charged weak interaction of the top quark is left-handed and flavor-changing, so expressed as
\begin{eqnarray}\label{eq1}
\frac{g_w}{2\sqrt{2}}V_{tf} \bar{t}(x)\gamma^\mu(1-\gamma_5)f(x) W_\mu(x),
\end{eqnarray}
where $f(x)$ stands for the fields of \textit{down}, \textit{strange} and \textit{bottom} quarks
and the weak coupling factor $g_w$ is related to the Fermi coupling constant as $g_w^2=4\sqrt{2} m_W^2 G_F$, 
while its neutral weak interaction is flavor-conserving and parity violating
\begin{eqnarray}
\frac{g_w}{4\cos\theta_w}\bar{t}(x)\gamma^\mu[(1-\frac{8}{3}\sin^2\theta_w)-\gamma_5]t(x) Z_\mu(x),
\end{eqnarray}
where $\theta_w$ is the weak mixing angle, so that $\sin^2\theta_w=0.23124$ \cite{Caso:1998tx}.
Its interaction with gluons is a vector-like coupling, involving an SU(3) 
generator ($T^a$) in the fundamental representation
\begin{eqnarray}\label{eq2}
g_s \bar{t}_i(x) \gamma^\mu T_{ij}^a t_j(x) G_\mu^a(x),
\end{eqnarray}
where $g_s$ is the strong coupling constant, $a=1,2,\cdots 8$ is the QCD color index so $Tr(T^a T^a)/3=C_F$.
The top interaction with photons is also simply vector-like as
\begin{eqnarray}
\frac{2}{3}e [\bar{t}(x)\gamma^\mu t(x) A_\mu(x)],
\end{eqnarray}
that is proportional to the top quark electric charge.\\ 
Finally, the interaction of the top quark with the Higgs field $h(x)$
is of the Yukawa type
\begin{eqnarray}\label{hasan}
y_t h(x)\bar{t}(x) t(x),
\end{eqnarray}
with a coupling constant $y_t=\sqrt{2}m_t/v$,
where $v$ is the Higgs vacuum expectation value. The Yukawa coupling is almost  $y_t\approx 1$ in the SM.\\
In many extensions of the SM such as minimal supersymmetric standard model (MSSM),
the Higgs sector of the SM is enlarged by considering an extra doublet of complex
 Higgs field \cite{Li:1990cp,Gunion:1984yn,Gunion}.
In \cite{MoosaviNejad:2011yp}, we studied the top decay in the general two Higgs doublet model (2HDM).\\
Moreover, beyond the interactions above, effective interactions such as for flavor-changing neutral currents occur
due to loop corrections. However, they are generally very small 
in comparison with those above. All these interactions could be modified in structure and strength 
by virtual effects due to new interactions associated with the physics beyond the SM. 
This is of interest to investigate that if the top quark, evidently, has a large coupling to the 
electroweak symmetry breaking sector. 
Therefore, it is so important to test these structures in detail, and indeed this is the thrust behind the field of top physics.\\
One proposed way to study the properties of top quarks is to consider
the scaled-energy distribution of outgoing hadrons.
 In next section we shall study this approach in detail, using
the GM-VFN scheme where the mass of b-quark is preserved from the beginning.

\section{Formalism}
We consider the decay process of polarized on-shell top quark at NLO, as
\begin{eqnarray}\label{eq3}
t(\uparrow)\rightarrow b+W^+(+g)\rightarrow B+X,
\end{eqnarray}
where, $X$ stands for the unobserved final-state particles. We wish to study the angular distribution
of the scaled-energy of B-hadrons by considering the contribution of bottom and gluon fragmentations
into the heavy meson B, so that the gluon contributes to the real radiation at NLO. 
To obtain this energy spectrum, we need to have the parton-level differential width
of the process (\ref{eq3}). The LO contribution results from $t(\uparrow)\rightarrow bW^+$.
We define the  partonic scaled-energy fraction $\textbf{x}_i=2p_i\cdot p_t/m_t^2$, where $p_i$
stands for the gluon or bottom quark momenta.
In the top quark rest frame where $p_t=(m_t, \vec{0})$, one has $\textbf{x}_i=2E_i/m_t$ where $E_i$
refers to the energy of outgoing partons; gluon or bottom at NLO.
By preserving the bottom quark mass, one has  $\textbf{x}_b^{max}= 1+b-\omega$ in which $b=m_b^2/m_t^2$
 and $\omega=m_W^2/m_t^2$.
As in \cite{Corcella:2001hz}, throughout this paper, we shall make use of the normalized energy fraction of partons as
\begin{eqnarray}\label{eq5}
x_i=\frac{\textbf{x}_i}{\textbf{x}_b^{max}}=\frac{2E_i}{m_t(1+b-\omega)},\quad (i=b,g).
\end{eqnarray}
The allowed values of $x_b$ and $x_g$ shall be discussed in Section \ref{sec5}.
We analyse the decay $t(\uparrow)\rightarrow bW^+$ in the rest frame of the top quark
where the 3-momentum of the b-quark points to the direction of the positive Z-axis.
For a polarized top quark, the general angular  distribution of  differential decay width $d\Gamma/dx$ is given
by
\begin{eqnarray}\label{eq6}
\frac{d^2\tilde\Gamma}{dx_i d\cos\theta_P}=\frac{1}{2}\bigg(\frac{d\tilde\Gamma^{unpol}}{dx_i}+
P\frac{d\tilde\Gamma^{pol}}{dx_i}\cos\theta_P\bigg).
\end{eqnarray}
This form clarifies  the correlations between the top 
decay products and the spin of the top quark.\\
 In (\ref{eq6}), $P$ is the magnitude of the top-quark polarization with $0\leq P\leq 1$, so that
$P=0$ corresponds to an unpolarized top quark and $P=1$ is for the $100\%$ polarization.
Here,  $\theta_P$  is defined as the polar angle between the top quark polarization vector $\vec{P}$
and the Z-axis (b-quark momentum direction).\\
In (\ref{eq6}), $d\tilde\Gamma^{unpol}/dx_i$ refers to the unpolarized differential widths 
which studied in \cite{Kniehl:2012mn}, both in ZM- and GM-VFN schemes. In following, we discuss 
the evaluation of the quantities $d\tilde\Gamma^{pol}/dx_i$ 
in the GM-VFN scheme.

\section{Parton-level results in the SM}
\subsection{Born term result}

Considering the charged weak interaction Lagrangian (\ref{eq1}), the dynamics of the
current-induced $t\rightarrow b$ transition is presented in the  tensor 
$H^{\mu\nu}\propto \left\langle t\right|J^{\nu\dag}\left|X_b\right\rangle \left\langle X_b\right|J^{\mu}\left|t\right\rangle$
in which the weak current is $J^\mu\propto V_{tb}\bar{t}(x)\gamma^\mu (1-\gamma_5)b(x)$, and
at the Born level and ${\cal O}(\alpha_s)$ one-loop contributions the intermediate state is $\left|X_b\right\rangle=\left|b\right\rangle$.
Also, this hadronic tensor  depends on the top spin $s_t$.
It is straightforward to compute the Born term contribution to the decay (\ref{eq3}).
In the top rest frame, the four-momentum of the bottom quark
is set to $p_b=(E_b; 0, 0, p_b)$ and the polarization four-vector of the top quark
is set as $s_t=P(0; \sin\theta_P\cos\phi_P, \sin\theta_P\sin\phi_P, \cos\theta_P)$.
Considering the general distribution (\ref{eq6}), the Born term helicity structure of partial rates, reads
\begin{eqnarray}\label{aziz}
\frac{d\tilde\Gamma^{\textbf{(0)}}}{d\cos\theta_{P}}&=&\frac{1}{2}\bigg\{\tilde\Gamma_A^{\textbf{(0)}}- 
P\tilde\Gamma_B^{\textbf{(0)}}\cos\theta_{P}\bigg\},
\end{eqnarray}
where, the LO polarized ($\tilde\Gamma_B^{\textbf{(0)}}$) and 
unpolarized ($\tilde\Gamma_A^{\textbf{(0)}}$) total decay widths  read  
\begin{eqnarray}
\tilde\Gamma_A^{\textbf{(0)}}&=&\frac{m_t \alpha Q}{4 \sin^2\theta_W}G_0,\nonumber\\
\tilde\Gamma_B^{\textbf{(0)}}&=&\frac{m_t \alpha Q^2}{4 \sin^2\theta_W}\frac{1-b-2\omega}{\omega}. 
\end{eqnarray}
Here, we used the following kinematic variables, in the notations of Ref.~\cite{Corcella:2001hz}
\begin{eqnarray}\label{variable}
&&S=\frac{1}{2}(1+b-\omega),\quad \beta=\frac{\sqrt{b}}{S}, \quad Q=S\sqrt{1-\beta^2},\nonumber\\
&&G_0=\frac{1}{2}(1+b-2\omega+\frac{(1-b)^2}{\omega}).
\end{eqnarray}
In the limit of vanishing bottom quark mass, the tree-level
decay widths converted to our results in \cite{Nejad:2013fba} and \cite{Kniehl:2012mn}, respectively.

\subsection{Virtual corrections and counterterms}
The QCD one-loop vertex corrections arise from the emission and absorption of the
virtual gluons, so an interaction Lagrangian  as in (\ref{eq2}) is needed to
calculate the virtual radiative corrections.  
Here, we adopt the on-shell mass renormalization scheme and use dimensional regularization
to regulate the ultraviolate (UV) and soft singularities which appear in one-loop corrections.
 For example, the UV-singularities
appear when the integration region of the internal momentum of the virtual gluon
goes to infinity. The singularities are regularized
by dimensional regularization in $D=4-2\epsilon$ space-time dimensions to
become single poles in $\epsilon$, so that $0<\epsilon\leq 1$.
In the massless case, all singularities are subtracted at factorization scale $\mu_F$
and absorbed into the bare FFs in accordance with the modified minimal subtraction ($\overline{MS}$) scheme, see \cite{Nejad:2013fba}.
In the massive case, all singularities are  automatically canceled after summing all radiative corrections up. \\
Considering the notations (\ref{variable}), the contribution
of virtual corrections into the doubly differential decay width (\ref{eq6}) is obtained as 
\begin{eqnarray}
\frac{d^2\tilde\Gamma^{vir}}{dx_b d\cos\theta_P}=
\frac{Q}{16\pi m_t}\bigg\{2 Re(M_0^\dagger M_{\rm 1-loop})\bigg\} \delta(1-x_b),
\end{eqnarray}
where $M_0$ stands for the Born term amplitude and
the renormalized amplitude  $M_{\rm 1-loop}$ refers to the virtual gluon corrections, presented in \cite{Nejad:2013fba}.
The virtual contributions include the counterterm and the one-loop vertex corrections.
The counterterm of the vertex contains the wave-function renormalization constants of the top
($\delta Z_t$) and the bottom quark ($\delta Z_b$). These constants can be found in \cite{Nejad:2013fba}.\\
The wave-function renormalization  and the one-loop vertex correction
contain the UV and infrared (IR) singularities so that all UV-divergences are canceled after 
summing all virtual corrections up and, from now on, we label the remaining IR-singularities
by $\epsilon$. Therefore, the virtual decay width is given by
\begin{eqnarray}\label{vir}
\frac{d^2\tilde\Gamma^{vir}}{dx_b d\cos\theta_P}=
\frac{1}{2}\Big(\frac{d\tilde\Gamma_A^{vir}}{dx_b} +P\frac{d\tilde\Gamma_B^{vir}}{dx_b}\cos\theta_P \Big)\delta(1-x_b),
\end{eqnarray}
where the unpolarized differential decay rate reads
\begin{eqnarray}
\frac{d\tilde\Gamma_A^{vir}}{dx_b}=
\tilde\Gamma_A^{(0)}\frac{C_F\alpha_s}{2\pi}
\Big\{\tilde{H}+\frac{6Q}{G_0}\ln\frac{S+Q}{\sqrt{b}}+\frac{3S(b-1)}{2\omega G_0}\ln b\Big\},
\end{eqnarray}
and the polarized one is expressed by
\begin{eqnarray}
\frac{d\tilde\Gamma_B^{vir}}{dx_b}=\tilde\Gamma_B^{(0)}\frac{C_F\alpha_s}{2\pi}
\Big\{\tilde{H}+\frac{3b-1}{2(1+3b-4S)}\ln b
+\frac{2(1+3S)\omega-2(1-b)}{Q(1+3b-4S)}\ln\frac{S+Q}{\sqrt{b}}\Big\},
\end{eqnarray}
where,
\begin{eqnarray}
\tilde{H}=2F\Big(-1+\frac{S}{Q}\ln\frac{S+Q}{\sqrt{b}}\Big)-
2\frac{Q+2S}{S}+\frac{2S}{Q}\ln(S+Q)\ln\frac{1-S-Q}{1-S+Q}-\ln b+
\frac{2S}{Q}\Big(Li_2\frac{2Q}{1+Q-S}-Li_2\frac{2Q}{S+Q-b}\Big).\nonumber\\
\end{eqnarray}
In the equations above, $C_F=(N_c^2-1)/(2N_c)=4/3$ is the color factor, $\gamma_E$ is the Euler constant, 
$Li_2(x)$ is the dilogarithmic function (or Spence function) and the term $F$ includes the IR-singularity ($\epsilon$) as
\begin{eqnarray}
F=\frac{1}{\epsilon}-\frac{Q}{S}-\gamma_E+\ln\frac{4\pi\mu_F^2}{m_t^2}-\ln\frac{S+Q}{\sqrt{b}}-\ln b.
\end{eqnarray}
Here, $\mu_F$ stands for the factorization scale which will be removed 
after summing all corrections up in the GM-VFN scheme.

\subsection{Real gluon corrections}
The ${\cal O}(\alpha_s)$ real graph contributions result from  the real gluon emissions 
from the bottom and top quarks, individually.
In the rest frame of a top quark decaying into a b-quark, a $W^+$ boson and a gluon, the outgoing particles
define an event plane. Relative to this plane one can, then, define the spin direction
of the polarized top quark. 
As in \cite{Nejad:2013fba}, here we apply  a specific helicity coordinate system where
the momenta of the b-quark and the $W^+$ boson  are defined as;
$\vec{p}_b || \hat{z} , (\vec{p}_W)_x\geq 0 $, and the polarization vector
of top quark is evaluated relative to the $\hat{z}$-axis. 
In the following, we explain a brief technical detail of our calculation for the NLO
radiative corrections to the tree-level decay rate of $t(\uparrow)\rightarrow bW$.

In \cite{Nejad:2013fba}, where we set the mass of b-quark to zero from the first,
the IR-singularities arised from the soft- and collinear gluon emissions.
Since, here, we preserve the mass of b-quark  then 
all IR-singularities arise from the soft real-gluon emission and
the collinear divergences would be absent.
As before, to regularize the IR-singularities we work in D-dimensions where
the real differential rate is given by
\begin{eqnarray}
d\tilde\Gamma^{real}=
\frac{\mu_F^{2(4-D)}}{2m_t}\left|M^{real}\right|^2
\prod_{i=b,g,W}\frac{d^{D-1}\vec{p_i}}{(2\pi)^{D-1}2E_i}(2\pi)^D\delta^D(p_t-\sum_{i=b, g, W}p_i).
\end{eqnarray}
To compute the differential rate $d\tilde{\Gamma}^{real}/dx_b$, we fix the b-quark momentum
and integrate over the gluon energy which ranges from
$E_g^{max}=m_t S (1-x_b)/(1-S x_b+S\sqrt{x_b^2-\beta^2})$ to 
$E_g^{min}=m_t S(1-x_b)/(1-S x_b-S\sqrt{x_b^2-\beta^2})$. 
In the GM-VFN scheme the real and virtual differential widths include the pole $\propto 1/\epsilon$,
which shall disappear in the total NLO result. 
Due to the radiation of a soft gluon ($E_g\rightarrow 0$)
in top decay, during integration over the phase space for the real
gluon radiation, terms of the form $(1-x_b)^{-1-2\epsilon}$ arise which are divergent when $x_b\rightarrow 1$.
Therefore, for a massive scheme ($m_b\neq 0$) where $\beta\leq x_b\leq 1$, we shall make use of the following expression \cite{Corcella:2001hz}
\begin{eqnarray}
\frac{(1-x_b)^{-1-2\epsilon}}{(x_b-\beta)^{-2\epsilon}}=-\frac{1}{2\epsilon}\delta(1-x_b)+
\frac{1}{(1-x_b)_+}+{\cal O}(\epsilon),
\end{eqnarray}
with the plus prescription defined as
\begin{eqnarray}
\int_\beta^1[g(x_b)]_{_{+}}h(x_b) dx_b=\int_\beta^1 g(x_b)[h(x_b)-h(1)]dx_b.
\end{eqnarray}
 
\subsection{Parton-level results for angular distribution of partial decay rates in FFN scheme}
Considering the tree-level, the real and virtual contributions, we present our
analytic expression for the angular distribution of the
partial decay rate in the FFN scheme. According to the Lee-Nauenberg theorem, 
after summing all corrections up the singularities cancel each other and the final
result is free of IR-singularities. Therefore, the complete NLO results read
\begin{eqnarray}\label{all}
\frac{d^2\tilde\Gamma}{dx_b d\cos\theta_P}=
\frac{1}{2}\Big(\frac{d\tilde\Gamma_{NLO}^{unpol}}{dx_b} +P\frac{d\tilde\Gamma_{NLO}^{pol}}{dx_b}\cos\theta_P \Big),
\end{eqnarray}
where $d\tilde\Gamma_{NLO}^{unpol}/dx_b$ is given in \cite{Corcella:2001hz,Kniehl:2012mn}, and 
$d\tilde\Gamma_{NLO}^{pol}/dx_b$ in the $\overline{MS}$ scheme is presented, for the first time, as
\begin{eqnarray} \label{alles}
\frac{1}{\tilde\Gamma_B^{(0)}}\frac{d\tilde\Gamma_{NLO}^{pol}}{dx_b}=&\delta(1-x_b)+\frac{C_F\alpha_s}{2\pi}
\Big\{\delta(1-x_b)\bigg[-4-
\frac{1-8S+9b}{2(1-b-2\omega)}\ln b+
\frac{1}{Q}\ln\frac{1+Q-S}{1-S-Q}
\bigg(1-b-2S\ln(S+Q)\bigg)+
\nonumber\\
&4\ln\frac{\sqrt{\omega}}{2S(1-\beta)}+\frac{4S}{Q}Li_2\frac{2Q}{1+Q-S}-\frac{4S}{Q}Li_2\frac{2Q}{S+Q-b}+
\frac{2S}{Q}\ln\frac{S+Q}{\sqrt{b}}\bigg(2\ln\frac{2S}{1+Q-S}-2\ln\frac{S+Q}{1-\beta}+
\nonumber\\
&1-\frac{b}{S}+\frac{1-b-\omega(1+3S)}{S(1-b-2\omega)}\bigg)\bigg]+
\frac{2(1-x_b)}{Q^2(1-b-2\omega)}\bigg[4S^3(x_b^2-2)
+\frac{2S\omega}{\sqrt{x_b^2-\beta^2}}\ln\phi_2(x_b)+2\frac{(1-b)S^2\omega}{1+b-2Sx_b}
\nonumber\\
&+bS(4-3Sx_b)+4S
-S^2(5+x_b-5b)+\frac{\phi_3(x_b)}{\sqrt{x_b^2-\beta^2}}\bigg(Sx_b(b+2\omega-1)-4(\omega+S^2x_b^2)\bigg)\bigg]+
\nonumber\\
&\frac{2}{Q^2(1-x_b)_+}\bigg[x_bS^2-b-\frac{2x_bS^3(1-x_b)^2}{1-b-2\omega}\bigg]
\bigg[x_b-2-\frac{x_b}{2\sqrt{x_b^2-\beta^2}}\bigg(\ln b+
2\frac{\phi_3(x_b)}{S}-2\ln\phi_1(x_b)\bigg)\bigg]\Big\},
\end{eqnarray}
where
\begin{eqnarray}
\phi_1(x_b)&=&\frac{S(x_b+\sqrt{x_b^2-\beta^2})-b}{1-S(x_b-\sqrt{x_b^2-\beta^2})},\nonumber\\
\phi_2(x_b)&=&\frac{S(x_b-\sqrt{x_b^2-\beta^2})-b}{S(x_b+\sqrt{x_b^2-\beta^2})-b},\nonumber\\
\phi_3(x_b)&=&S(\sqrt{x_b^2-\beta^2}-artanh\frac{\sqrt{x_b^2-\beta^2}}{x_b}).
\end{eqnarray}
Since the B-mesons can be also produced through the fragmentation of emitted real gluons, then
to obtain the most accurate result for the energy spectrum of mesons one needs
the doubly differential distribution $d^2\tilde\Gamma/(dx_g d\cos\theta_P)$ where the scaled-variable $x_g$
is defined in (\ref{eq5}).
As we will show in Fig.~\ref{fig1}, the contribution of the gluon fragmentation into the B-meson
is negative and leads to a significant reduction in size in the threshold region, so that this contribution would be
important at a low energy of the observed meson.\\
To get the  $d^2\tilde\Gamma/(dx_g d\cos\theta_P)$,
one has to integrate over the b-quark energy by fixing the gluon momentum in the phase-space
so that the b-quark energy ranges as $ m_t S\phi_{-}(x_g)\leq E_b \leq m_t S\phi_{+}(x_g)$, where
\begin{eqnarray} 
\phi_\pm(x_g)=\frac{1-x_g}{1-2 S x_g}\big[1-Sx_g\pm Sx_g\sqrt{1-\frac{(1-2Sx_g)\beta^2}{(1-x_g)^2}}\big].
\end{eqnarray}
The $d\tilde\Gamma_{NLO}^{unpol}/dx_g$ in the FFN scheme is given in \cite{Kniehl:2012mn}, and by
considering the following notations
\begin{eqnarray} 
A_1=&x_gS^3(1+(1-x_g)^2)+\frac{8S^2(1-S)(1-x_g)}{3(1+3b-4S)}x_g-\frac{S^2}{3}(4x_g^2+2x_g+3)+
S(\frac{x_g}{2}+\frac{2}{3})+\frac{1}{6}(1+3b-4S)(1-3Sx_g)-\frac{1}{6},
\nonumber\\
&A_2=\frac{S^3x_g(1-x_g)^3}{[(1-x_g)^2-\beta^2]^{\frac{3}{2}}}\Big[1+(1-x_g)^2-
\frac{4+3x_g(x_g-2)}{(1-x_g)^2}\beta^2+\frac{2+x_g(Sx_g-2)}{(1-x_g)^3}\beta^4\Big],
\end{eqnarray}
and $A_3=\frac{2}{3}S\bigg(3+Sx_g^2+(3S-4)x_g-8\frac{(1-S)(1-Sx_g)}{1+3b-4S}x_g\bigg)$,
the polarized contribution reads 
\begin{eqnarray} \label{hame}
\frac{1}{\tilde\Gamma_B^{(0)}}\frac{d\tilde\Gamma_{NLO}^{pol}}{dx_g}&=&
\frac{C_F\alpha_s}{4\pi Q^2x_g^2}
\Big\{2\frac{A_1}{S}\ln\frac{\phi_{+}+\sqrt{\phi_{+}^2-\beta^2}}{\phi_{-}+\sqrt{\phi_{-}^2-\beta^2}}+2\frac{A_2}{S}
\bigg[\ln\frac{(1-2Sx_g)(1-x_g-\phi_+)^2}{bx_g^2}
\nonumber\\
&&+\ln\frac{\beta^2+(x_g-1)\phi_{-}-\sqrt{((1-x_g)^2-\beta^2)(\phi_{-}^2-\beta^2)}}
{\beta^2+(x_g-1)\phi_{+}-\sqrt{((1-x_g)^2-\beta^2)(\phi_+^2-\beta^2)}}\bigg]+
A_3(\sqrt{\phi_{+}^2-\beta^2}-\sqrt{\phi_{-}^2-\beta^2})
\nonumber\\
&&+\frac{2S(1-2Sx_g)}{\beta^2-(1-x_g)^2}(1-x_g-\beta^2)\bigg[\sqrt{\phi_{-}^2-\beta^2}(x_g-1+\phi_+)-
\sqrt{\phi_{+}^2-\beta^2}(x_g-1+\phi_{-})\bigg]
\nonumber\\
&&+S(1-Sx_g)\bigg[\phi_{-}\sqrt{\phi_{-}^2-\beta^2}-\phi_{+}\sqrt{\phi_{+}^2-\beta^2}\bigg]
\Big\}.
\end{eqnarray}

\subsection{General-mass variable-flavor-number scheme}

In \cite{Nejad:2013fba}, for obtaining  the parton-level results for angular distribution of partial decay rates 
we used the ZM-VFN scheme, where $m_b=0$ was put right from the beginning
and all  collinear singularities were absorbed into the bare FFs according to the
$\overline{MS}$ scheme. This approach renormalizes the FFs and produces finite terms of the form $(\alpha_s/\pi)\ln(\mu_F^2/m_t^2)$ in the 
partial decay rates $d\hat\Gamma/dx_a$, which are rendered perturbatively small
by choosing $\mu_F={\cal O}(m_t)$. In this scheme, the b-quark mass $m_b$ just sets the initial scale
$\mu_F^{ini}={\cal O}(m_b)$ of the DGLAP evolution equations, 
where ansaetze for the $z$-dependences of the FFs $D_a(z, \mu_F^{ini})$ are injected
by some proposed models \cite{jm}.
The DGLAP evolution from $\mu_F^{ini}$ to a higher scale $\mu_F$ then
effectively resums the problematic logarithms $(\alpha_s/\pi)\ln(m_t^2/m_b^2)$ of the FFN scheme,
however, all information on the $m_b$-dependence of $d\hat\Gamma/dx_a$ is wasted.

The GM-VFN scheme provides an ideal theoretical framework to study the effects of heavy quark masses,
so it combines the virtues of the ZM-VFN and FFN schemes and, at the same time, avoids their flaws.
In the GM-VFN scheme, the perturbative fragmentation functions enter the formalism via subtraction
terms for the hard scattering decay rates, so that the actual FFs are truly nonperturbative
and may be assumed to have some smooth forms which can be specified through global data fits.
In opposition with the FFN scheme, the GM-VFN scheme also accommodates FFs for light quarks and gluons, 
as in the ZM-VFN scheme. In our present work, the GM-VFNs is applied to resum the
large logarithms in $m_b$ and to retain the entire
nonlogarithmic $m_b$-dependence at the same time. This is reached by introducing
convenient subtraction terms in the  ${\cal O}(\alpha_s)$ FFN expressions for $d\tilde\Gamma/dx_i$, so
that the ${\cal O}(\alpha_s)$ ZM-VFN results are exactly recovered in the limit $m_b/m_t\rightarrow 0$.
These subtraction terms are universal and so are the FFs in the FFN scheme, as is guaranteed by 
Collin's hard-scattering factorization theorem \cite{Collins:1998rz}.

As explained above, the GM-VFN results for the 
angular decay distributions are obtained by matching the FFN results (\ref{alles},\ref{hame})
to the ZM-VFN ones \cite{Nejad:2013fba} by subtraction,  as
\begin{eqnarray}\label{mahsa}
\frac{1}{\Gamma_0}\frac{d\Gamma}{dx_i}\Big|_{GM-VFN}=
\frac{1}{\Gamma_0}\frac{d\tilde\Gamma}{dx_i}\Big|_{FFN}-\frac{1}{\Gamma_0}\frac{d\Gamma}{dx_i}\Big|_{Sub},
\end{eqnarray}
where the subtraction terms are obtained as
\begin{eqnarray}
\frac{1}{\Gamma_0}\frac{d\Gamma}{dx_i}\Big|_{Sub}=
\lim_{m_b\rightarrow 0}\frac{1}{\Gamma_0}\frac{d\tilde\Gamma}{dx_i}\Big|_{FFN}-
\frac{1}{\Gamma_0}\frac{d\hat\Gamma}{dx_i}\Big|_{_{ZM-VFN}}.
\end{eqnarray}
Taking the limit $m_b\rightarrow 0$ in (\ref{alles}) and (\ref{hame}), we recover the results presented in
\cite{Nejad:2013fba} up to the terms
\begin{eqnarray}\label{ayda}
\frac{1}{\Gamma_0}\frac{d\Gamma}{dx_b}\Big|_{Sub}=
\frac{\alpha_s(\mu_R)}{2\pi}C_F\Big\{\frac{1+x_b^2}{1-x_b}
\Big[\ln\frac{\mu_F^2}{m_b^2}-2\ln(1-x_b)-1\Big]\Big\}_+,
\end{eqnarray}
and,
\begin{eqnarray}\label{mona}
\frac{1}{\Gamma_0}\frac{d\Gamma}{dx_g}\Big|_{Sub}=
\frac{\alpha_s(\mu_R)}{2\pi}C_F\frac{1+(1-x_g)^2}{x_g}
\Big(\ln\frac{\mu_F^2}{m_b^2}-2\ln x_g-1\Big).
\end{eqnarray}
As we have already shown in \cite{Kniehl:2012mn}, for the unpolarized top decay in the SM, i.e. $t\rightarrow bW^+$, and also
for the top decay in the theories beyond the SM including the two Higgs doublet where $t\rightarrow bH^+$ \cite{MoosaviNejad:2011yp},
Eq.~(\ref{ayda}) coincides with the perturbative FF of the transition $b\rightarrow b$.
This is in agreement with the Collin's factorization theorem which 
guarantees that the subtraction terms are universal. Thus the results presented
in (\ref{ayda}) and (\ref{mona}) ensure the correctness  of our calculations shown in (\ref{alles},\ref{hame}).

\section{Hadron mass effects and Hadron level results}
\label{sec5}

Our main purpose is to obtain the scaled-energy ($x_B$) distribution 
of bottom-flavored hadrons (B) inclusively produced
in  polarized top decays at NLO.
Here, the  scaled-energy fraction of B-hadrons is defined as $x_B=2E_B/(m_t(1+b-\omega))$, as in (\ref{eq5}).
In \cite{Nejad:2013fba}, to obtain the partial decay width of the process (\ref{eq3})
 in the ZM-VFN scheme, we used
the Collin's factorization theorem \cite{Collins:1998rz}. According to this theorem, 
the energy distribution of B-hadrons might be expressed as the convolution of the partonic 
hard scattering  decay rates $d\Gamma/dx_a$,  with the
nonperturbative FFs which describe the transition $a\rightarrow B$, as
\begin{eqnarray}\label{convolute}
\frac{d\Gamma}{dx_B}=\sum_{a=b, g}\int_{x_a^{min}}^{x_a^{max}}
\frac{dx_a}{x_a}\frac{d\hat\Gamma}{dx_a}(\mu_R, \mu_F)D_a^B(\frac{x_B}{x_a},\mu_F).
\end{eqnarray}
Here, $\mu_F$ and $\mu_R$ are the factorization and the renormalization scales, respectively.
The $\mu_R$ is associated with the renormalization of the strong coupling constant and
a choice often applied is $\mu_R=\mu_F$.

In the massless (or ZM-VFN) scheme where one sets $m_b=0$, the b-quark, gluon and B-hadron (with the mass $m_B$)
have energies $0\leq(E_b, E_g)\leq(m_t^2-m_W^2)/(2m_t)$ and $m_B\leq E_B\leq(m_t^2+m_B^2-m_W^2)/(2m_t)$, respectively.

As  we demonstrated in \cite{Kniehl:2012mn}, the relation (\ref{convolute}) is 
convenient for the case  $m_b=0=m_B$. To calculate the $d\Gamma/dx_B$ when
passing from the ZM-VFN scheme to the GM-VFN scheme
by taking into account the finite-$m_B$ corrections, one should apply the following improved relation
\begin{eqnarray}\label{convolute2}
\frac{d\Gamma}{dx_B}=\frac{1}{\sqrt{x_B^2-\rho_B^2}}\sum_{a=b, g}\int_{x_a^{min}}^{x_a^{max}}
dx_a z\frac{d\Gamma}{dx_a}\Big|_{_{GM-VFN}}D_a^B(z,\mu_F),
\end{eqnarray}
where $(d\Gamma/dx_a)_{_{_{GM-VFN}}}$ is given in (\ref{mahsa}), and 
\begin{eqnarray}
z=\frac{x_B+\sqrt{x_B^2-\rho_B^2}}{x_a+\sqrt{x_a^2-\rho_a^2}},
\end{eqnarray}
with $\rho_i=m_i/E_b^{max}(i=b, g, B)$, so $E_b^{max}=(m_t^2+m_b^2-m_W^2)/(2m_t)$.
 Now, the kinematically allowed scaling-variables
are 
\begin{eqnarray} 
&&\frac{1}{2}(x_B+\sqrt{x_B^2-\rho_B^2}+\frac{\rho_b^2}{x_B+\sqrt{x_B^2-\rho_B^2}})\leq x_b\leq 1,\nonumber\\
&&\frac{1}{2}(x_B+\sqrt{x_B^2-\rho_B^2})\leq x_g\leq \frac{m_t^2-(m_b+m_W)^2}{m_t^2+m_b^2-m_W^2},\nonumber\\
&&\rho_B\leq x_B\leq \frac{m_t^2+m_B^2-m_W^2}{m_t^2+m_b^2-m_W^2}\quad \textrm{for b-quark transition},\nonumber\\
&&\rho_B\leq x_B\leq \frac{m_t^2+m_B^2-(m_b+m_W)^2}{m_t^2+m_b^2-m_W^2}\quad \textrm{for gluon transition}.
\end{eqnarray}
Clearly, if $m_b=0$ and $m_B=0$ are put ($z\rightarrow x_B/x_a$), 
then (\ref{convolute}) and (\ref{convolute2}) coincide by reproducing the familiar
factorization formula of the massless parton model.

\section{Numerical analysis}

\begin{figure}
\begin{center}
\includegraphics[width=0.5\linewidth,bb=97 40 780 680]{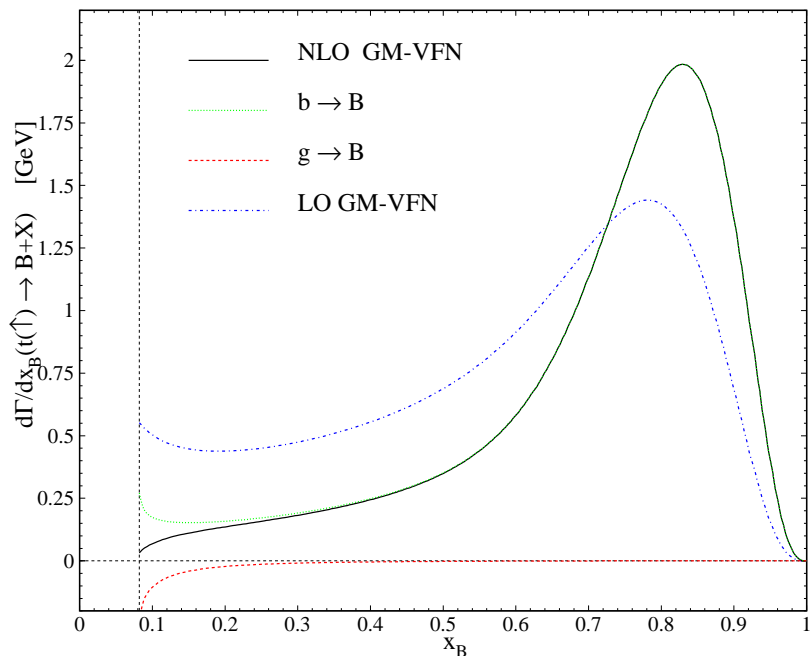}
\caption{\label{fig1}%
$d\Gamma(t(\uparrow)\rightarrow B+X)/dx_B$ as a function of $x_B$ at NLO in the GM-VFN ($m_b\neq 0$)
scheme including finite-$m_B$ corrections by using Eq.(\ref{convolute2}).
The NLO prediction (solid line) is compared to the LO one (dot-dashed line) and broken up 
into the contributions due to $b\rightarrow B$ (dotted line) and $g\rightarrow B$ (dashed line) fragmentation.}
\end{center}
\end{figure}

By having the necessary tools, we now turn our attention to the phenomenological predictions
of  the B-hadron energy spectrum in polarized top decays, considering
the effects of bottom quark and B-hadron masses.
From \cite{Nakamura:2010zzi}, we adopt the input parameter values 
$m_W=80.399$~GeV, $m_t=172.98$~GeV,$m_b=4.78$~GeV, $m_B=5.279$~GeV, $\sin^2\theta_W=0.2312$
and we evaluate $\alpha_s^{(n_f)}(\mu_R)$ at NLO in the $\overline{MS}$ scheme, using
\begin{eqnarray}
\alpha_s^{(n_f)}(\mu_R)=
\frac{1}{b_0\log[\mu_R^2/(\Lambda^{(n_f)})^2]}\bigg\{1-\frac{b_1\log(\log[\mu_R^2/(\Lambda^{(n_f)})^2])}
{b_0^2\log[\mu_R^2/(\Lambda^{(n_f)})^2]}\bigg\},
\end{eqnarray}
where, $n_f$ is  the number of active quark flavors and
\begin{eqnarray}
b_0=\frac{33-2n_f}{12\pi}, \quad b_1=\frac{153-19n_f}{24\pi^2}.
\end{eqnarray}
Considering $n_f=5$, we adopt the asymptotic scale parameter $\Lambda_{\overline{MS}}^{(5)}=231$~MeV,  adjusted
such that $\alpha_s^{(5)}(m_Z)=0.1184$ for $m_Z=91.1876$~GeV \cite{Nakamura:2010zzi}. To include the B-meson
and the b-quark masses, we apply Eq.~(\ref{convolute2}) in which for the transitions $(b, g)\rightarrow B$,
from \cite{Kniehl:2008zza} we employ the related realistic and nonperturbative FFs.
In \cite{Kniehl:2008zza}, a power model as $D_b^B(z, \mu_0)=Nz^\alpha(1-z)^\beta$
is used for the transition $b\rightarrow B$  at the initial scale $\mu_F=4.5$~GeV of fragmentation, while 
the FFs of gluon and light quarks are set to zero at the starting scale and 
are evolved to higher scales via the DGLAP equations \cite{dglap}.
The fit parameters are obtained  at NLO in the ZM-VFN scheme through
a global fit to $e^+e^-$ annihilation data from the 
ALEPH  and OPAL collaborations at CERN LEP1 and by SLD at SLAC SLC and the
results are $N=4684.1,\alpha=16.87$ and $\beta=2.628$. \\
In Fig.~\ref{fig1}, our predictions for the scaled-energy ($x_B$)
distribution of  B-hadrons are shown by considering  the corresponding quantity 
$d\Gamma(t(\uparrow)\rightarrow BX)/dx_B$ in the GM-VFN scheme. 
For this studying, we considered 
the size of the NLO corrections,
by comparing the LO (dot-dashed line) and NLO (solid line) results, and the relative importance
of the $b\rightarrow B$ (dotted line) and $g\rightarrow B$ (dashed line) fragmentation channels at NLO.
To compare the size of the NLO corrections at the parton level, we  evaluate the LO result
using the same NLO FFs. 
As is seen, the $g\rightarrow B$ contribution into the NLO energy spectrum of the 
B-meson is negative and appreciable only
in the low-$x_B$ region and for higher values of $x_B$, the NLO result is
practically exhausted by the $b\rightarrow B$ contribution, as expected in \cite{Corcella:2001hz}.
In fact, the contribution of the gluon is evaluated to see where
it contributes to $d\Gamma/dx_B$ and can not be discriminated in the meson spectrum  as an experimental quantity.
In the scaled-energy of mesons, all contributions
including the bottom quark, gluon and light quarks contribute.\\
From Fig.~\ref{fig1}, it is also seen that the NLO corrections lead to a significant
enhancement of the partial decay width in the peak region and above, by as much as $30\%$,
at the expense of a depletion in the lower-$x_B$ range.
Moreover, the peak position is shifted towards higher values of $x_B$.

\begin{figure}
	\begin{center}
		\includegraphics[width=0.5\linewidth,bb=97 40 780 680]{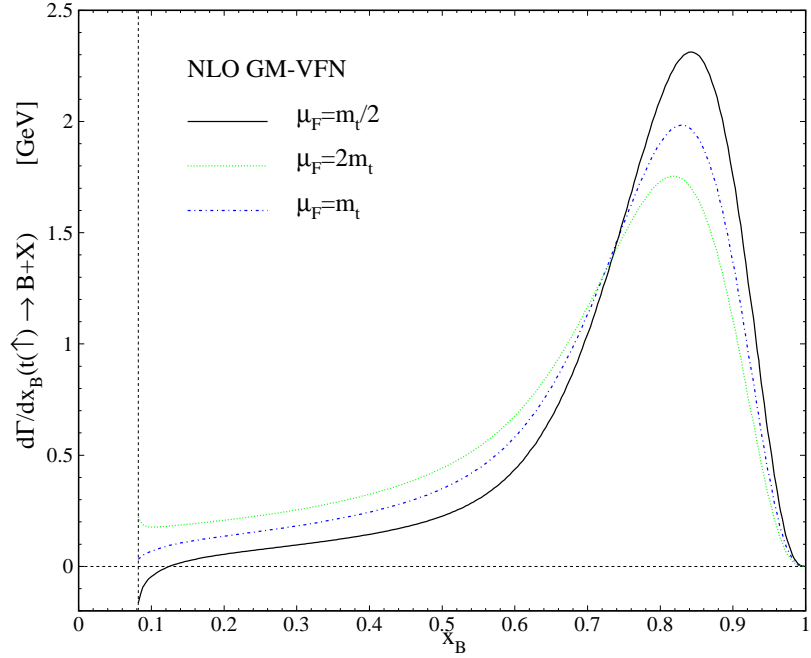}
		\caption{\label{band}%
			$d\Gamma(t(\uparrow)\rightarrow B+X)/dx_B$ at NLO in the GM-VFN ($m_b\neq 0$)
			 scheme including finite-$m_B$ corrections, considering 
			the factorization scales $\mu_F=m_t$ (dot-dashed line), $\mu_F=m_t/2$ (solid line) and $\mu_F=2m_t$ (dotted line).}
	\end{center}
\end{figure}

In (\ref{convolute}), the factorization ($\mu_F$) and the renormalization ($\mu_R$) scales are arbitrary 
and, in principle, one can use two different values for them. However, a choice
often made consists of setting  $\mu_R=\mu_F$ and we shall adopt this convention for most of the
results which we shall show.
In Fig.~\ref{band}, we show the dependence of the meson energy spectrum  
on the factorization scales by considering $\mu_F=m_t$ (dot-dashed line), $\mu_F=m_t/2$ (solid line) and $\mu_F=2m_t$ (dotted line).
In \cite{Corcella:2001hz}, the dependence of the $x_b$ spectrum 
on the factorization scales $\mu_{0F}$ and $\mu_F$ are studied in detail. 
Their results show that the $x_b$ dependence
on the initial scales $\mu_{0F}$ is small when one resums soft logarithms in the initial condition of the
b-quark perturbative FF. According to their results, as a whole conclusion, one can states
that resumming soft logarithms yields a reduction of the theoretical uncertainty, as 
the dependence on factorization scales is indeed an estimate of effects of
higher order contributions which we have been neglecting.

\begin{figure}
\begin{center}
\includegraphics[width=0.50\linewidth,bb=97 40 780 680]{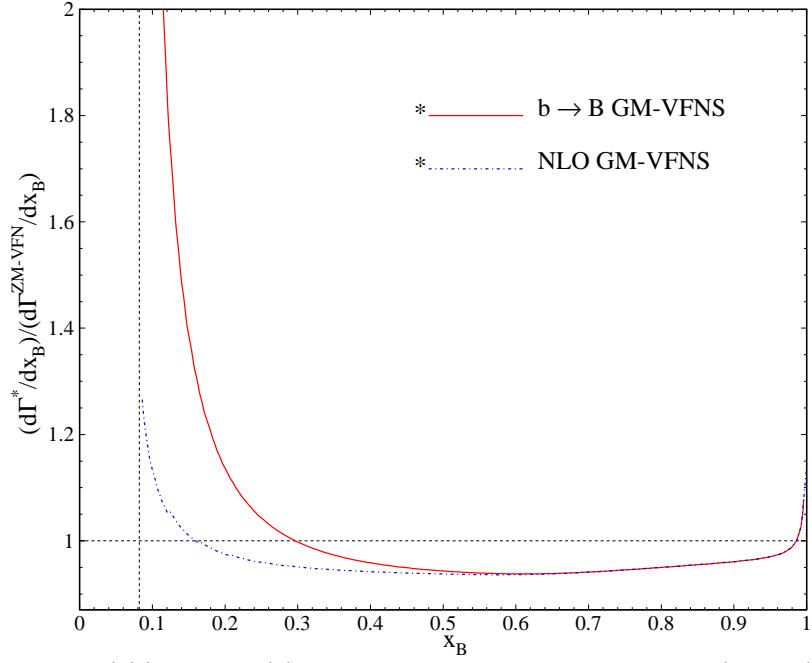}
\caption{\label{fig3}%
The result for the $d\Gamma(t(\uparrow)\rightarrow B+X)/dx_B$ is shown at NLO in the GM-VFN ($m_b\neq 0$)
scheme including finite-$m_B$ corrections (dot-dashed line). For comparison, also
the GM-VFN result excluding $g\rightarrow B$ fragmentation is shown (solid line).
All results are normalized to the ZM-VFN one for $m_B=0$, including $g\rightarrow B$ fragmentation.}
\end{center}
\end{figure}

\begin{figure}
	\begin{center}
		\includegraphics[width=0.50\linewidth,bb=97 40 780 680]{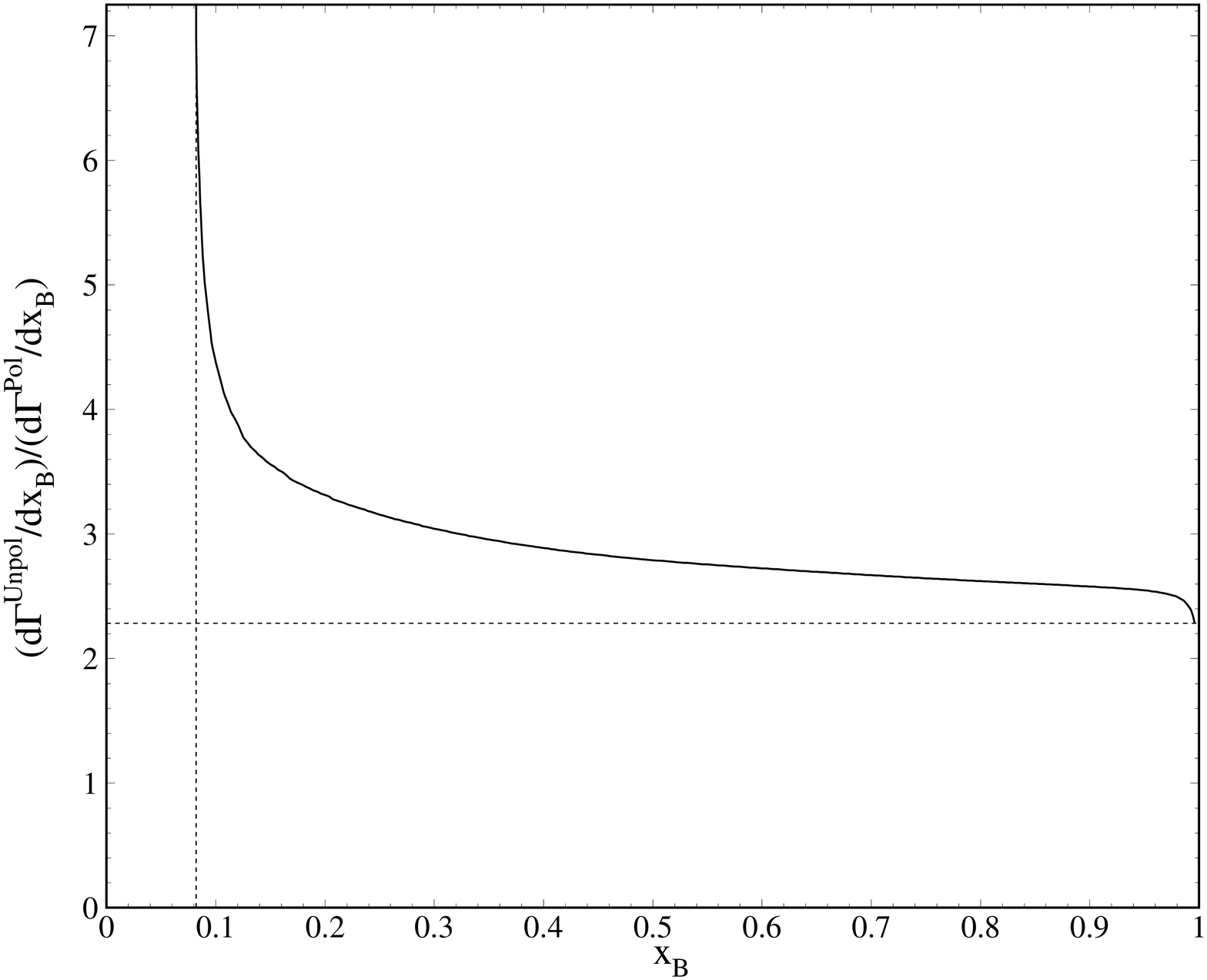}
		\caption{\label{ratio}%
			The ratio of the {\it unpolarised} and {\it polarised} results 
			for the $d\Gamma(t\rightarrow B+X)/dx_B$ at NLO in the GM-VFN ($m_b\neq 0$)
			scheme including finite-$m_B$ corrections.}
	\end{center}
\end{figure}

In Fig.~\ref{fig3}, for a quantitative comparison of our previous predictions \cite{Nejad:2013fba} at NLO in the ZM-VFN scheme, we consider
the GM-VFN result ($m_b\neq0$) including finite-$m_B$ corrections
without $g\rightarrow B$ fragmentation (solid line) and the full 
GM-VFN result (dot-dashed line) using Eq.(\ref{convolute2}), both 
normalized to the full ZM-VFN  result for $m_B=0$. It is observed that the omission 
of $g\rightarrow B$ fragmentation causes an excess by a factor of up to $2$ close to threshold, 
while the finite-$m_b$ and $m_B$ corrections  amount less. Note that, our most reliable prediction for the energy spectrum of B-meson is made 
at NLO in the GM-VFN scheme by including finite-$m_B$ corrections.
Relative to our previous work the improvements in our new work is twofold.
First, the finite-$m_B$ corrections 
are responsible for the appearance of the threshold at $x_B=\rho_B=0.083$, and second,
the finite-$m_b$ corrections lead to a  moderate reduction
in size throughout the whole $x_B$ range allowed, specially in the peak range.\\
In Fig.~\ref{ratio}, to study the angular dependence of energy distributions 
we plot the ratio of the  NLO unpolarized \cite{Kniehl:2012mn} and  polarized results 
for the $d\Gamma(t\rightarrow B+X)/dx_B$  in the GM-VFN ($m_b\neq 0$) scheme including finite-$m_B$ corrections. 
\section{Conclusions}
\label{sec:four}

Studying the fundamental properties of the top
quark is an object of interest in theoretical and experimental particle physics.
Among other things, the LHC is a superlative top factory which allows
one to study the top characteristics in great detail, if also
the theoretical descriptions and simulations are of commensurate quality.
In particular, the LHC will allow for the 
study of the dominant decay mode $t\rightarrow BW^++ X$ with unprecedented precision in the long run.
As an application, these studies will  enable us to deepen our conception of the nonperturbative 
aspects of B-hadron formation by hadronization and to pin down the $b\rightarrow B$ and $g\rightarrow B$ fragmentation functions.
The key quantity for this purpose is
the $x_B$ distribution $d\Gamma/dx_B$ of
$t\rightarrow B+X$. Therefore, the distributions in the scaled B-hadron energy $x_B$
through the polarized (\ref{eq3}) or unpolarized  \cite{Kniehl:2012mn} top decays
 are of particular interest at the LHC.
In this context, recently, the local CMS group at the CERN LHC started to work on a determination of the top quark mass
from a detailed study of the B-meson decays.\\
The top quark decays rapidly so that has no enough time
to hadronize and then passes on its full spin information to its decay products.
This allows one to study the top spin state using the angular distributions of its decay products,
so that, in this work we studied  
the ${\cal O}(\alpha_s)$ spin-dependent energy spectrum of hadrons produced from polarized top quark decays.
For this, we studied the observable $d\Gamma/dx_B$ at NLO in the GM-VFN scheme \cite{Kniehl:2008zza}. 
 This allowed us to investigate, for the first time, finite-$m_b$ corrections to the $d\Gamma/dx_B$.
 We also analyzed the size of finite-$m_B$ effects. 
Specifically, our analysis is supposed to enhance  our previous results presented in \cite{Nejad:2013fba} 
by retaining all nonlogarithmic $m_b$ terms of the result in the FFN scheme.
These studies are mandatory in order to fully exploit the enormous 
statistics of the LHC data to be taken in the long run for a high-precision 
determination of the top-quark properties.\\
Comparing future measurements of the polarized  width $d\Gamma(t(\uparrow)\rightarrow BW^++X)/dx_B$ at the LHC with our NLO predictions, 
one will be also able to test the universality and scaling violations of the B-hadron FFs. 
These measurements of the $x_B$ distributions will ultimately be the primary source of information on the B-hadron FFs.\\
Our formalism elaborated here is also applicable to the production of hadron species other than B-hadrons, such as pions, kaons and
protons, etc.,  using the $b, g\rightarrow \pi/K/P$ FFs presented in our recent paper \cite{Soleymaninia:2013cxa},
relying on their universality and scaling violations.


\begin{thebibliography}{25}

\bibitem{Abe:1995hr}
  F.~Abe {\it et al.} [CDF Collaboration],
  Phys.\ Rev.\ Lett.\  {\bf 74} (1995) 2626
  [hep-ex/9503002].

\bibitem{Abachi:1995iq}
  S.~Abachi {\it et al.} [D0 Collaboration],
  Phys.\ Rev.\ Lett.\  {\bf 74} (1995) 2632
  [hep-ex/9503003].

\bibitem{Abazov:2014dpa}
  V.~M.~Abazov {\it et al.} [D0 Collaboration],
  Phys.\ Rev.\ Lett.\  {\bf 113} (2014) 032002
  [arXiv:1405.1756 [hep-ex]].
	
\bibitem{Abazov:2012vd}
  V.~M.~Abazov {\it et al.} [D0 Collaboration],
  Phys.\ Rev.\ D {\bf 85} (2012) 091104
  [arXiv:1201.4156 [hep-ex]].

\bibitem{Aaltonen:2013kna}
  T.~A.~Aaltonen {\it et al.} [CDF Collaboration],
  Phys.\ Rev.\ Lett.\  {\bf 111} (2013) 20,  202001
  [arXiv:1308.4050 [hep-ex]].

\bibitem{delDuca:2015gca}
  V.~del Duca and E.~Laenen,
  Int.\ J.\ Mod.\ Phys.\ A {\bf 30} (2015) 35,  1530063
  [arXiv:1510.06690 [hep-ph]].


\bibitem{Cabibbo:1963yz}
  N.~Cabibbo,
  Phys.\ Rev.\ Lett.\  {\bf 10} (1963) 531.
  M.~Kobayashi and T.~Maskawa,
  Prog.\ Theor.\ Phys.\  {\bf 49} (1973) 652.
	
\bibitem{Chetyrkin:1999ju}
  K.~G.~Chetyrkin, R.~Harlander, T.~Seidensticker and M.~Steinhauser,
  Phys.\ Rev.\ D {\bf 60} (1999) 114015
  [hep-ph/9906273].


\bibitem{Nejad:2013fba}
  S.~M.~M.~Nejad,
  Phys.\ Rev.\ D {\bf 88} (2013) 9,  094011
  [arXiv:1310.5686 [hep-ph]];
  S.~M.~Moosavi Nejad and M.~Balali,
  Phys.\ Rev.\ D {\bf 90} (2014) 11,  114017
  [arXiv:1409.1389 [hep-ph]].


\bibitem{Corcella}
G.~Corcella and F.~Mescia,
Eur.\ Phys.\ J.\ C {\bf 65} (2010) 171;
G.~Corcella and F.~Mescia,
Eur.\ Phys.\ J.\ C {\bf 68} (2010) 687 (Erratum);
S.~Biswas, K.~Melnikov and M.~Schulze,
JHEP {\bf 1008} (2010) 048,
[arXiv:1006.0910 [hep-ph]].

\bibitem{Corcella:2001hz}
G.~Corcella and A.~D.~Mitov,
Nucl.\ Phys.\ B {\bf 623} (2002) 247
[hep-ph/0110319].

\bibitem{Collins:1998rz} 
  J.~C.~Collins,
  Phys.\ Rev.\ D {\bf 58}, 094002 (1998)
  [hep-ph/9806259].
	

	
\bibitem{dglap}
  V.~N.~Gribov and L.~N.~Lipatov,
  Sov.\ J.\ Nucl.\ Phys.\  {\bf 15} (1972) 438
   [Yad.\ Fiz.\  {\bf 15} (1972) 781];
  G.~Altarelli and G.~Parisi,
  Nucl.\ Phys.\ B {\bf 126} (1977) 298.
  Y.~L.~Dokshitzer,
  Sov.\ Phys.\ JETP {\bf 46} (1977) 641
   [Zh.\ Eksp.\ Teor.\ Fiz.\  {\bf 73} (1977) 1216].

\bibitem{Caso:1998tx}
  C.~Caso {\it et al.} [Particle Data Group Collaboration],
  Eur.\ Phys.\ J.\ C {\bf 3} (1998) 1.
 	
 	
 \bibitem{Li:1990cp}
 C.~S.~Li and T.~C.~Yuan,
 Phys.\ Rev.\ D {\bf 42} (1990) 3088
 [Phys.\ Rev.\ D {\bf 47} (1993) 2156].
 	
 
 \bibitem{Gunion:1984yn}
 J.~F.~Gunion and H.~E.~Haber,
 Nucl.\ Phys.\ B {\bf 272} (1986) 1
 [Nucl.\ Phys.\ B {\bf 402} (1993) 567].
 	
\bibitem{Gunion}
J.~F.~Gunion, H.~Haber, G.~Kane, and S.~Dawson, {\it The
Higgs Hunter's Guide} (Addison-Wesley, Reading, MAA,
1990), and refrences therein.


\bibitem{MoosaviNejad:2011yp}
  S.~M.~Moosavi Nejad,
  Phys.\ Rev.\ D {\bf 85} (2012) 054010
  [arXiv:1110.1601 [hep-ph]];
  S.~M.~Moosavi Nejad,
  Eur.\ Phys.\ J.\ C {\bf 72} (2012) 2224
  [arXiv:1205.6139 [hep-ph]].
 

	

\bibitem{Kniehl:2012mn}
  B.~A.~Kniehl, G.~Kramer and S.~M.~Moosavi Nejad,
  Nucl.\ Phys.\ B {\bf 862} (2012) 720
  [arXiv:1205.2528 [hep-ph]].
	

\bibitem{jm}
  J.~Binnewies, B.~A.~Kniehl and G.~Kramer,
  Phys.\ Rev.\ D {\bf 58} (1998) 034016
  [hep-ph/9802231].
	

\bibitem{Nakamura:2010zzi}
  K.~Nakamura {\it et al.} [Particle Data Group Collaboration],
  J.\ Phys.\ G {\bf 37} (2010) 075021.
	
	
\bibitem{Kniehl:2008zza}
  B.~A.~Kniehl, G.~Kramer, I.~Schienbein and H.~Spiesberger,
  Phys.\ Rev.\ D {\bf 77} (2008) 014011
  [arXiv:0705.4392 [hep-ph]].

\bibitem{Soleymaninia:2013cxa}
  M.~Soleymaninia, A.~N.~Khorramian, S.~M.~Moosavi Nejad and F.~Arbabifar,
  Phys.\ Rev.\ D {\bf 88} (2013) 5,  054019
   [Phys.\ Rev.\ D {\bf 89} (2014) 3,  039901]
  [arXiv:1306.1612 [hep-ph]];
  S.~M.~M.~Nejad, M.~Soleymaninia and A.~Maktoubian,
  arXiv:1512.01855 [hep-ph].

 
\end{thebibliography}
\end{document}